# A Review on Energy, Environmental, and Sustainability Implications of Connected and Automated Vehicles

Morteza Taiebat,[†,‡] Austin L. Brown,[§] Hannah R. Safford,[∇] Shen Qu,[†] and Ming Xu*,[†,‡]

[†]School for Environment and Sustainability, University of Michigan, Ann Arbor, Michigan 48109, United States

[‡]Department of Civil and Environmental Engineering, University of Michigan, Ann Arbor, Michigan 48109, United States

[§]Policy Institute for Energy, Environment, and the Economy, University of California, Davis, California 95616, United States

[∇]Department of Civil & Environmental Engineering, University of California, Davis, California 95616, United States

Supporting Information

**ABSTRACT:** Connected and automated vehicles (CAVs) are poised to reshape transportation and mobility by replacing humans as the driver and service provider. While the primary stated motivation for vehicle automation is to improve safety and convenience of road mobility, this transformation also provides a valuable opportunity to improve vehicle energy efficiency and reduce emissions in the transportation sector. Progress in vehicle efficiency and functionality, however, does not necessarily translate to net positive environmental outcomes. Here, we examine the interactions between CAV technology and the environment at four levels of increasing complexity: *vehicle*, *transportation system*, *urban system*, and *society*. We find that environmental impacts come from CAV-facilitated transformations at all four levels, rather than from CAV technology directly. We anticipate net positive environmental impacts at the vehicle, transportation system, and urban system levels, but expect greater vehicle utilization and shifts in travel patterns at the society level to offset some of these benefits. Focusing on the vehicle-level improvements associated with CAV technology is likely to yield excessively optimistic estimates of environmental benefits. Future research and policy efforts should strive to clarify the extent and possible synergetic effects from a systems level to envisage and address concerns regarding the short- and long-term sustainable adoption of CAV technology.

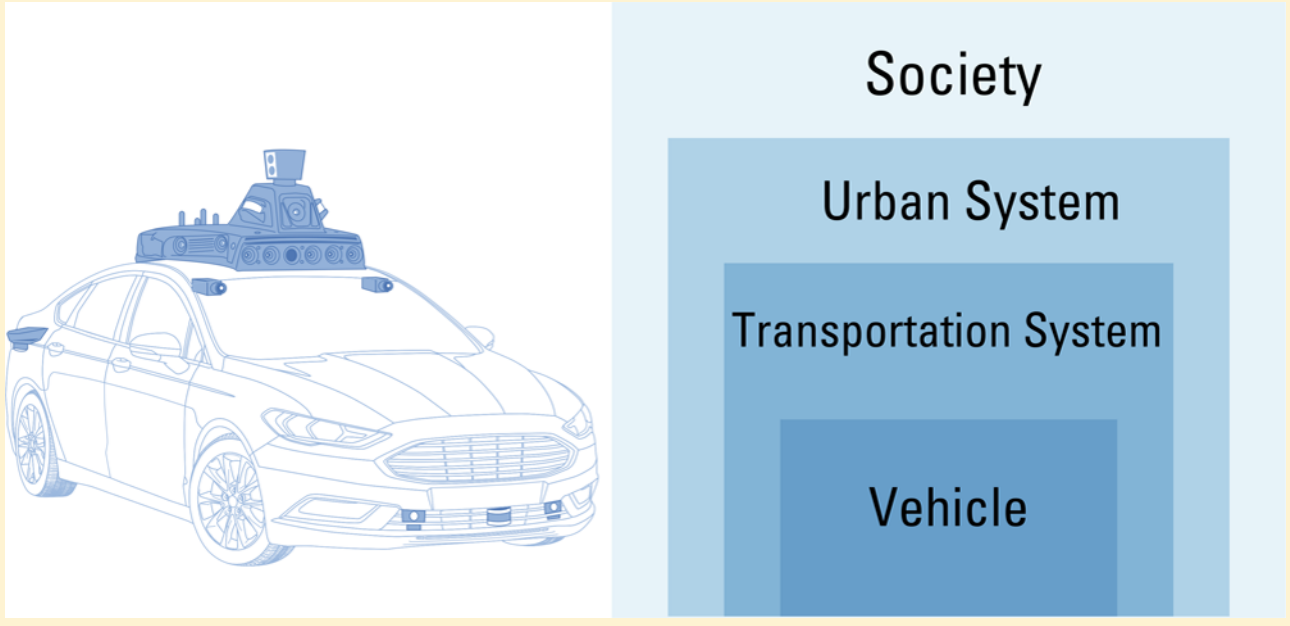

## 1. INTRODUCTION

The fuel-based transportation system holds considerable influence over human interactions with the environment. Transportation directly generated over 7 gigatons of carbon dioxide equivalent ($GtCO_2$ equiv) greenhouse gas (GHG) emissions worldwide in 2010 or 23% of total global energy-related GHG emissions.[1] Annual transportation GHG emissions are increasing at a faster rate than emissions from any other sector (i.e., power, industry, agriculture, residential, or commercial). With income rising and infrastructure expanding around the world, transportation demand is expected to increase dramatically in the coming years. Annual transportation sector emissions are expected to double by 2050.[1]

In the U.S., the transportation sector was the largest source of GHG emissions in 2016, accounting for 28.5% of total national energy-related GHG emissions, according to the U.S. Environmental Protection Agency (EPA).[2] The most recent data from the U.S. Energy Information Administration (EIA) also shows that carbon dioxide ($CO_2$) emissions from the U.S. transportation sector (1893 million metric tons or MMt) surpassed $CO_2$ emissions from the electric power sector (1803 MMt) from October 2015 through September 2016.[3] This is the first time that transportation-sector $CO_2$ emissions have regularly exceeded $CO_2$ emissions from the electric power sector since the late 1970s on a 12-month rolling basis. This trend is likely to continue if growth in renewable energy lowers fossil fuel-based electricity generation, leading to continued reduction of power sector emissions.

Within the transportation sector, road-based travel is responsible for the largest share of $CO_2$ emissions, GHG emissions, and energy use compared to other modes of transportation such as aviation, rail, and marine. Passenger cars, light-duty trucks (including sport utility vehicles, pickup trucks, and minivans), and freight trucks emitted 41.6%, 18.0%, and 22.9%, respectively, of total U.S. transportation-sector GHG emissions in 2016.[2] Given that emissions from the transportation sector increased more in absolute terms than emissions from any other sector from 1990−2016, transportation emissions must be a key focus of mitigation efforts. Strategic development and deployment of new technologies to curb the environmental impacts of road-based travel can therefore go a long way toward alleviating the environmental impacts of the transportation sector overall. One example with considerable potential to reduce emissions from road-based travel is connected and automated vehicle (CAV) technology.

Vehicle connectivity and automation are separate technologies that could exist independent of each other, but entail strong complementary attributes. Connectivity refers to a vehicle's capacity to exchange information with other vehicles and infrastructure. This capacity can be realized through vehicle-to-vehicle (V2V), vehicle-to-infrastructure (V2I), and other cooperative communications networks. Vehicle connectivity is a key enabler of vehicle automation. Vehicle automation refers to any instance in which control of a vehicle capability normally overseen by a human driver is ceded to a computer. Examples of automation commonly seen in vehicles on the market today include cruise control, adaptive cruise control, active lane-keep assist, and automatic emergency braking. A fully automated vehicle can navigate itself by sensing and interacting with the driving environment to reach its destination without human intervention.[4−6]

It is worth noting that the terms "autonomous" and "automated" are often used interchangeably in the literature, but merit distinction. The former (a subset of the latter) refers to a vehicle capable of navigating without direct input from a human driver and self-driving is possible with limited or no communication with other vehicles or infrastructure, while the latter indicates broader classes of vehicle automation. In this article, the term "CAV technology" refers to vehicle technology with high levels of automation, as well as connectivity capabilities. These two facets of CAV technology are expected to develop in concert.

The Society of Automotive Engineers (SAE) International's J3016 taxonomy classifies vehicle automation by level of driver intervention and/or attentiveness required for operation.[7] To avoid redundancy and confusion, the U.S. National Highway Traffic Safety Administration (NHTSA) agreed to adopt the SAE's categorization, instead of relying on vehicle capabilities.[8] In 2016, the NHTSA proposed mandating V2V connectivity capability on all new cars and light-duty trucks, citing significant potential safety benefits.[9] On September 12, 2017, the U.S. Department of Transportation released updated federal guidelines for the deployment of highly automated vehicle technologies.[10] These guidelines focus on road safety performance and mobility services, without addressing environmental impacts.

The primary purpose of CAV technology is to increase transportation safety and provide better mobility services.[10] However, vehicle connectivity and automation will also inevitably and significantly change the environmental profile of the transportation sector.[11−15] A growing body of literature has examined the possible environmental implications of CAVs, and has found large uncertainty based in part on the shortage of real-world data for CAV operations.[16] CAV technology could facilitate either dramatic decarbonization of transportation or equally dramatic increases in transportation-sector emissions. The net environmental impacts of CAV technology depend on lawmaking and decisions at the international, federal, state, and local levels. With the transition to automated road transportation still in its infancy, there is an opportunity to work proactively to ensure that CAV technology develops sustainably. A forward-looking perspective is needed to properly design, plan, and develop a CAV system that provides both better mobility service and better environmental outcomes.

This article is intended to foster understanding and discussion of the likely and potential environmental implications of CAV technologies by reviewing existing studies and identifying key research needs. We define environmental impacts broadly in this paper, including not only downstream emissions and wastes, but also upstream resource and energy demands. We also discuss some socioeconomic aspects of CAV adoption that are associated with energy and the environment. Our review includes some environmental impacts that could be realized through vehicle automation alone, but most impacts require automation in conjunction with connectivity. For simplicity, we attribute all impacts to CAV technology. The article is organized as follows. We begin by developing a holistic framework for analyzing different levels of interactions between CAVs and the environment (section 2). We then survey the quantitative results of relevant studies and critically evaluate the key assumptions and conclusions of each (section 3). Finally, we identify knowledge gaps and offer recommendations for future research (section 4).

## 2. LEVELS OF INTERACTIONS BETWEEN CAVS AND THE ENVIRONMENT

CAV technology interacts with the environment at different scales and levels of complexity. We define four levels of interactions between CAVs and the environment—the *vehicle* level, *transportation system* level, *urban system* level, and *society* level—as illustrated in Figure 1. Interactions generally increase in complexity from the vehicle level to society level and may stem from CAV technology directly or CAV-facilitated effects.

The most direct and well-studied interactions occur at the vehicle level. At this level, connectivity and automation physically alter vehicle design and operation. At the transportation system level, CAV technology can drastically change how vehicles interact with each other in the driving environment. At the urban system level, CAV-based transportation interacts with a wide range of infrastructure in the urban environment such as roads, power grid, and buildings, thereby altering how urban systems utilize resources and energy and generate emissions and waste. Finally, how the public perceives and how the government regulates CAVs can have profound effects at the society level.

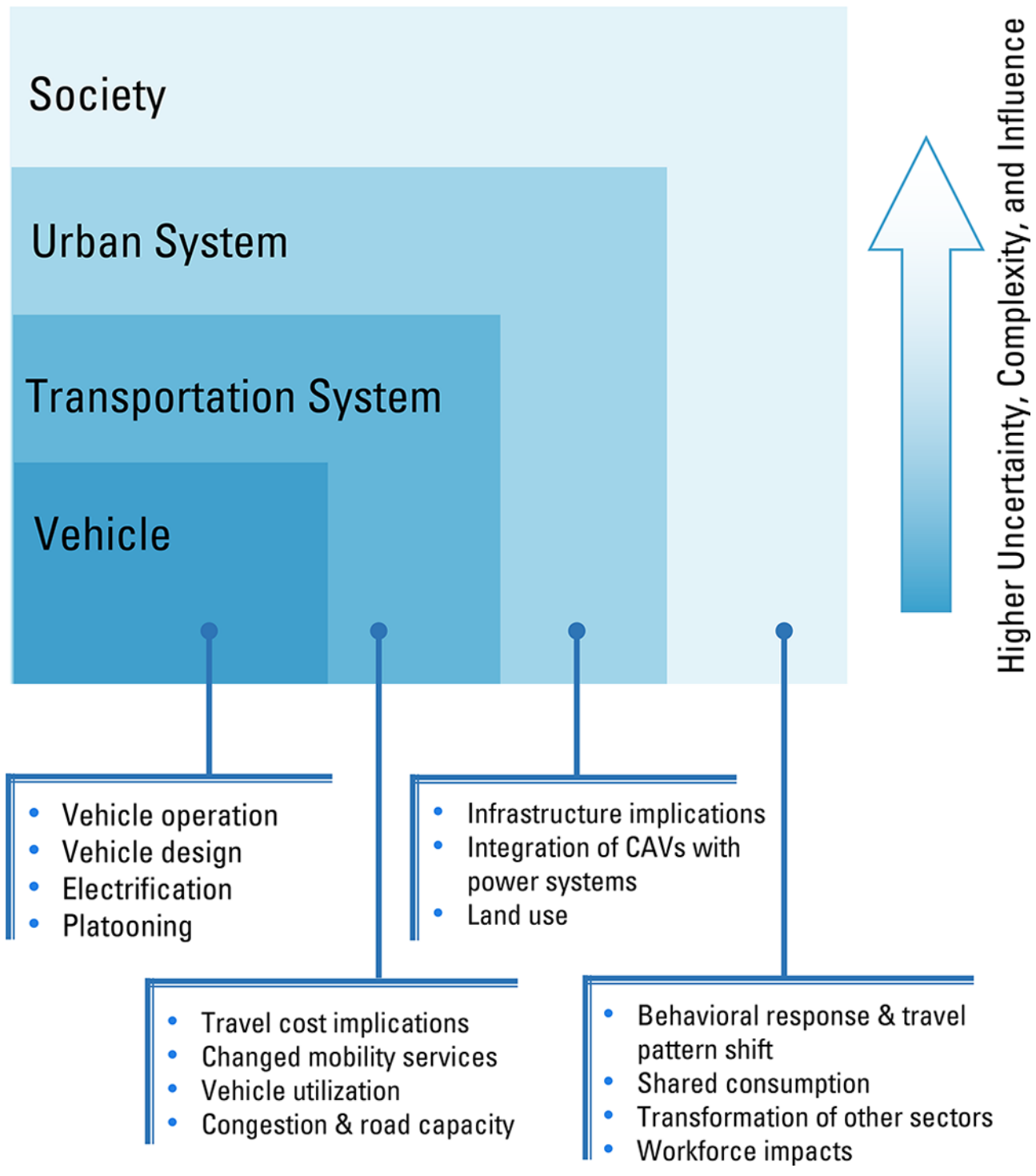

**Figure 1.** Levels of interactions between CAVs and the environment and corresponding major influence mechanisms.

Table 1. Summary of Key Environmental Impacts at Each Level of CAV-Environment Interaction

| | Major Influencing Mechanisms | Positive Impacts | Negative Impacts | Sources |
|---|---|---|---|---|
| **Vehicle** | •vehicle operation<br>•vehicle design<br>•electrification<br>•platooning | •higher energy efficiency<br>  •optimal driving cycle<br>  •eco-routing<br>  •reduce cold starts<br>  •less idling<br>  •less speed fluctuations<br>  •powertrain downsizing<br>  •self-parking<br>•safety-enabled vehicle light-weighting<br>•vehicle right-sizing<br>•complementary electrification benefits<br>•platooning | •faster highway speeds<br>•additional ICT equipment needs for navigation and communication<br>•aerodynamic shape alteration<br>•higher auxiliary power requirement | 4, 12−14, 16−22 |
| **Transportation System** | •travel-cost implications<br>•changed mobility services<br>•vehicle utilization<br>•congestion and road capacity | •greatly reduced human labor costs<br>•promotion of shared mobility<br>•integration with mass transit<br>•fleet downsizing<br>•increased effective roadway capacity<br>•decongestion<br>•fewer crashes and less accident-related traffic<br>•syncing with traffic lights | •higher vehicle utilization rate<br>•more frequent and longer trips result in greater VMT<br>•more unoccupied travel (for parking, between trips, etc.)<br>•congestion increases due to induced travel<br>•competition with mass transit | 12−14, 16, 17, 19, 23−32 |
| **Urban System** | •infrastructure implications<br>•integration of CAVs with power systems<br>•land use | •changes in land-use patterns<br>•reduced need for parking infrastructure<br>•integration with power systems through vehicle electrification<br>•reduced need for highway lighting and traffic signals | •increased urban sprawl<br>•need for large, energy-intensive data centers | 14, 24, 33−36 |
| **Society** | •behavior response and travel pattern shift<br>•shared consumption<br>•transformation of other sectors<br>•workforce impacts | •promotion of shared consumption<br>•spillover effects to other sectors | •induced travel demand and rebound effect<br>•transportation modal shift (e.g., rail/aviation to road travel)<br>•gradual unemployment and job displacement | 16, 17, 25, 37−42 |

Generally, higher-level interactions will have farther-reaching implications despite often receiving less attention (Table 1). Higher-level interactions are also more difficult to quantify and are associated with greater uncertainty. Many important questions at high levels are beyond the scope of quantitative or predictive modeling and must instead be addressed qualitatively. Because research focusing on CAV environmental implications is just emerging in recent years, a large body of literature is in the form of reports and white papers. In order to make this review as comprehensive as possible, our analysis is based on not only peer-reviewed studies but also reputable reports and documents containing consensus quantitative results. Key sources are classified based on scope in Table 2.

## 3. ENVIRONMENTAL IMPACTS OF CAV AT EACH SYSTEM LEVEL

### 3.1. Vehicle Level.

At this level, we consider the direct environmental effects of CAV technology on a per vehicle basis. These effects can also manifest in fleets. Many studies have focused on the vehicle level and show that individual CAVs are generally more energy efficient and generate less emissions than conventional vehicles.[12,13,16] These benefits at the vehicle level can be attributed to four major factors: operation, electrification, design, and platooning.

*3.1.1. Vehicle Operation.* A number of references discuss the potential for vehicle automation to improve car-centric energy efficiency by optimizing vehicle operation: that is, by maximizing the operation of vehicles at the most efficient mode.[6,14,19,30] Efficient driving broadly translates into improved fuel economy, reduced energy consumption, and abated tailpipe emissions. Higher driving efficiency can be achieved in CAVs through a variety of mechanisms, including optimal driving cycle, dynamic eco-routing, less idling, reducing cold starts, trip smoothing, and speed harmonization.[12,14,28,29,59] These mechanisms are discussed below.

Different human drivers in identical situations make different real-time decisions, often leading to suboptimal results.[5] In CAVs, eliminating heterogeneity between drivers and improving driving decision-making helps optimize the driving cycle. Barth and Boriboonsomsin reported that, even when drivers remain "in the loop" of vehicle operation (i.e., at a level of involvement less than conventional driving but one that falls short of full automation), providing dynamic feedback to drivers results in up to 20% fuel savings and decreased $CO_2$ emissions without a significant increase in travel time.[30] The information gathered from vehicle connectivity also enables optimal route selecting, widely known as dynamic eco-routing.[19,30,63] Gonder et al. estimated the potential energy savings of eco-routing in a Chevy Bolt at around 5%.[49] Trip smoothing and speed harmonization are other practices that aim to minimize repeated braking-acceleration cycles through intelligent speed adaption, smooth starts, fewer speed fluctuations, and eliminating unnecessary full stops.

**Table 2. Classification of relevant CAV studies by scope**

| Study[a] | Vehicle | Transp. sys. | Urban sys. | Society |
|---|---|---|---|---|
| Alonso-Mora et al.[24 b] | | √ | | |
| Anderson et al.[6] | √ | √ | √ | √ |
| Auld et al.[25 b] | | √ | | |
| Bansal and Kockelman[38 b] | | | | √ |
| Barth et al.[19] | √ | √ | | |
| Bauer et al.[43 b] | √ | √ | | |
| Brown et al.[14] | √ | √ | √ | √ |
| Chen et al.[31 b] | | √ | √ | |
| Chen et al.[44 b] | √ | | | |
| Childress et al.[17 b] | | √ | √ | √ |
| Crayton and Meier[45 b] | | | | √ |
| Fagnant and Kockelman[29 b] | √ | √ | | |
| Fox-Penner et al.[46 b] | | | √ | |
| Fulton et al.[47] | √ | √ | | |
| Gawron et al.[48 b] | √ | | | |
| Gonder et al.[49 b] | √ | | | |
| Greenblat and Shaheen[32] | √ | √ | √ | √ |
| Greenblatt and Saxena[13 b] | √ | √ | √ | |
| Harper et al.[40 b] | | | | √ |
| Heard et al.[50 b] | | | | √ |
| Kang et al.[23 b] | √ | √ | √ | |
| Kolosz and Grant-Muller[35 b] | | √ | √ | |
| König and Neumayr[51 b] | | | | √ |
| Kyriakidis et al.[41 b] | | | | √ |
| Lavrenz and Gkritza[52 b] | √ | | √ | |
| Li et al.[36 b] | √ | | √ | |
| Liu et al.[53] | √ | | | |
| Lu et al.[26 b] | | √ | √ | |
| Malikopoulos et al.[54b] | √ | √ | | |
| Mersky and Samaras[18 b] | √ | | | |
| Moorthy et al.[55 b] | | √ | | |
| Prakash et al.[56] | √ | | | |
| Rios-Torres and Malikopoulos[20] | √ | √ | | |
| Stephens et al.[16] | √ | √ | √ | √ |
| Stern et al.[28 b] | √ | | | |
| Wadud[57 b] | | √ | | √ |
| Wadud et al.[12 b] | √ | √ | √ | √ |
| Wang et al.[58 b] | √ | | | |
| Wu et al.[59 b] | √ | | | |
| Zakharenko[60 b] | | √ | √ | |
| Zhang et al.[61 b] | | | √ | |
| Zhang et al.[62 b] | | √ | | |

[a]Sorted alphabetically based on first author. [b]Publication in a peer-reviewed journal.

CAV technology substantially facilitates and amplifies these practices. Wu et al. estimated that partial automation in conjunction with connectivity can reduce fuel use by 5−7% compared to human driving when automation enables vehicles to closely follow recommended speed profiles.[59] At the fleet level, cooperative communications between vehicles can further reduce energy use, with up to 13% fuel savings and 12% reductions in $CO_2$ emissions reported in experiments.[19] Prakash et al. suggested that 12−17% reduction in fuel use can be achieved when a CAV is trailing a lead vehicle with the specific objective of minimizing accelerations and decelerations.[56] On the basis of experiments, Stern et al. found that introducing even a single CAV into traffic dampens stop-and-go patterns, results in up to 40% reductions in total traffic fuel consumption.[28] Rios-Torres and Malikopoulos developed a simulation framework for mixed traffic (CAVs interacting with human-driven vehicles) and reported that the fuel-consumption benefits of CAVs increase with higher CAV penetration.[20] Chen at al. suggested a wider range of changes in fuel consumption (between −45% to +30%) that would result from transitioning from conventional to CAV fleets at the U.S. national level.[44]

Less idling and fewer cold starts can help reduce energy waste and mitigate emissions. Cold starts are a major contributor to a number of criteria air pollutants from the transportation sector, including volatile organic compounds (VOCs), $NO_x$, and CO.[19] Simulations demonstrated fewer cold starts for shared automated taxis.[29] In such vehicles, since no aggressive acceleration is needed, powertrains can also be downsized. This is especially relevant for automated shared mobility services in urban areas where more energy use is due to acceleration rather than from high-speed wind resistance.[12] Self-parking features also save time and limit braking-acceleration cycles, reducing energy intensity by approximately 4%.[14]

On the other hand, some attributes of CAVs may result in more energy consumption. Radar, sensors, network communications, and high-speed Internet connectivity require higher auxiliary power from vehicles, which manifests as greater power draw and consequently higher energy consumption.[64] Energy demands for connectivity components, sensing, and computing equipment can significantly alter the overall energy efficiency of CAVs.[48] Additionally, improved safety in CAVs may induce higher highway speeds. Since aerodynamic drag forces increase quadratically with speed, higher highway speeds result in higher fuel consumption above a certain threshold.[19] For instance, a speed increase from 70 to 80 mile per hour (MPH) is reported to increase average energy use by 13.9% per mile.[65] Wadud et al. and Brown et al. suggested that typical driving at above-optimal speeds tends to decrease overall fuel economy by 5−22%.[12,14] This decrease may offset—and indeed, overwhelm—increases in engine efficiency. It is conceivable that improved safety in CAVs could enable relaxation of speed limits for roadways where vehicles are currently restricted to below-optimal speeds, resulting in some energy savings. This point received less attention in the literature.

The extent to which CAV-related increases in vehicle energy consumption will offset gains in energy efficiency is unclear. CAV technology could lead to substantial net improvements in fuel economy and emissions reduction if the negative effects are minimized and the positive realized. Mersky and Samaras raised the question of how to test and measure fuel efficiency of CAVs by updating EPA rating tests.[18] They developed a method for testing fuel economy of CAVs using the existing EPA test procedure and showed that fuel economy differences for the CAV tests range from −3% to +5% compared to the current EPA testing procedure.

*3.1.2. Electrification.* Many studies examining the environmental externalities of vehicle electrification have found that electric vehicles (EVs) usually improve environmental outcomes and remove local pollution from urban cores.[66,67] The specific environmental impacts of EVs are largely determined by when cars are charged and where and how chargers are integrated into the electric grid. Emissions from power generation for EVs might in some cases be higher than tailpipes emissions from vehicles with internal combustion engines. However, moving emissions from a large number of individual vehicle tailpipes to a few centralized power plants is likely to reduce emission mitigation costs, improve energy efficiency, and help integrate renewable energy in power generation.[66] Offer et al. demonstrated that plug-in hybrid electric vehicles (PHEVs) and battery electric vehicles (BEVs) have much lower life-cycle costs and emissions compared to fuel cells or internal combustion engines vehicles.[68] Despite potential benefits, the actual environmental impacts of EVs are affected by many factors, such as unregulated charging, vehicle-to-grid (V2G) communications, charge speed, and the degree to which users overcome range anxiety. The effects of these factors remain uncertain and require more research.

CAV technology can provide a strong complement to EV technology, potentially solving some of the challenges of EV development.[14] In electric CAVs, on-board energy management strategies can be explicitly designed and implemented to take advantage of synergies between electrification and automation. For instance, an electric CAV could optimize route selection and driving cycle to reduce battery draining, maximize energy recovery via regenerative braking, and extend the battery life.

CAVs can also mitigate the range restriction of EVs by matching appropriately ranged vehicles to individual trips,[31] and take advantage of the energy and environmental benefits brought by vehicle electrification. Offer argued that even if electric CAVs substantially increase vehicle utilization, they will have a large positive impact on transport decarbonization and will curb global GHG emissions by improving the economics of electrification.[21] Shared automated electric vehicles (SAEVs) magnify benefits by orders of magnitude.[46] Greenblatt and Saxena suggested that electric automated taxis can reduce per-mile GHG emissions by more than 90% compared to using conventional vehicles for daily travel.[13] Bauer et al. simulated the operation of SAEVs in NYC, and found that under the current power-grid mix, SAEV fleet would generate 73% fewer GHG emissions and consume 58% less energy than a nonelectrified automated fleet.[43]

*3.1.3. Vehicle Design.* The size and weight of a vehicle have direct impacts on the vehicle's fuel economy, and consequently on its overall environmental performance. The composition of the vehicle body indirectly influences the life-cycle environmental impacts of the vehicle via resource and energy requirements associated with the supply chain. CAV engineering is expected to enable a number of efficiency-improving design practices, such as vehicle right-sizing and safety-enabled vehicle light-weighting. On the other hand, more carbon-intensive materials are needed in CAVs, which could increase overall per-vehicle weight as well. Differences in CAV design strategies among automakers and the evolution of Evolution of design design over time add uncertainties to analysis of CAV-related environmental impacts.

*3.1.3.a. Vehicle Light-Weighting.* A number of recent studies have addressed the life-cycle environmental impacts of vehicle light-weighting using alternative materials. Several report that each 10% reduction in vehicle weight yields on average a direct fuel economy improvement of 6−8%.[14,69]

In a highly connected and automated vehicle system, transportation safety can be significantly improved by eliminating human errors in driving. As a result, once CAVs make up the vast majority of on-road active vehicles, crashworthiness of vehicles becomes less crucial, and vehicles can become smaller with less safety equipment. Safety features contributed to 7.7% of total vehicle weight in an average new

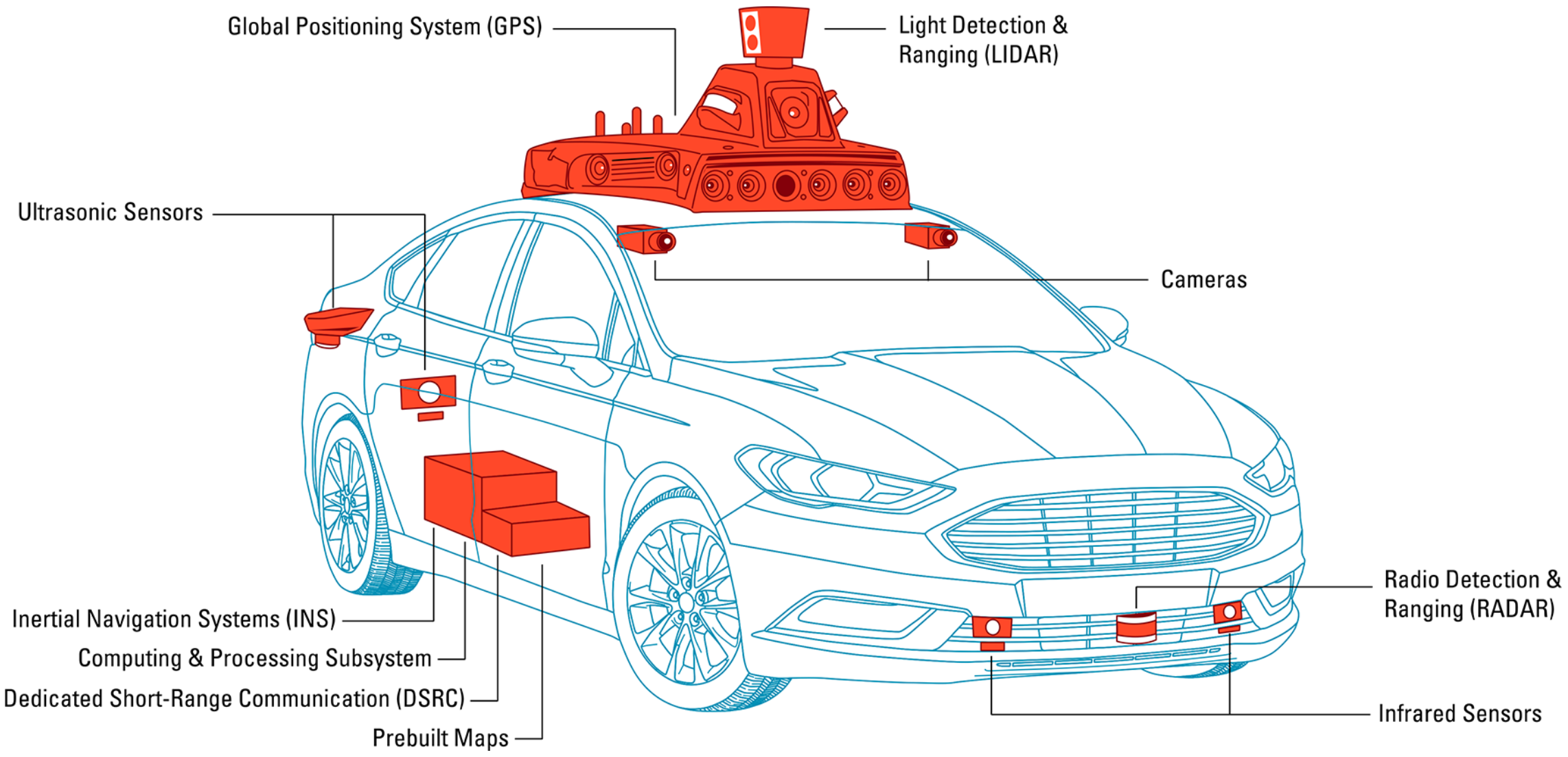

**Figure 2.** Key technologies and additional ICT devices in a generic CAV for navigation and communication. This figure is a generalized model based on components and subsystems described in the literature.[6,73] Actual engineering designs will vary among automakers and vehicle models, and future designs are likely to change as CAV engineering advances. Additional information about these components are provided in SI (S1).

U.S. vehicle in 2011.[12] If these features could be safely removed, an estimated 4.6−6.2% improvement in fuel economy could be realized.[14] Moreover, environmental impacts associated with the life-cycle of the eliminated vehicle safety features could also be avoided.

Reduced safety equipment in CAVs also leads to more optimal and smaller powertrains, further improving fuel economy. Wadud et al. suggested “de-emphasized performance” as another potential option that would further downsize the powertrain of CAVs and save up to 5% of fuel consumption.[12] Conventional vehicles typically have power capabilities far in excess of their average power requirements to satisfy occasional high-performance demands, such as freeway merging. The ability of CAVs to smooth speed profiles, coupled with the high potential of CAVs to serve in shared mobility services, means that peak power demand could be significantly reduced.

*3.1.3.b. Vehicle Right-Sizing.* Another opportunity that could be realized from widespread use of CAVs is vehicle “right-sizing”. According to 2017 National Household Travel Survey, single- and double-occupant vehicle trips respectively accounted for 58% and 25% of total annual vehicle-miles-traveled (VMT) in passenger trips made in the U.S., and the average occupancy of light-duty vehicles was just 1.67 passengers.[70] There is significant potential for vehicle size optimization by matching specific vehicles to specific trips to avoid wasted capacity and thus associated environmental impacts. In the case of automated taxis or shared automated vehicles (SAVs), a vehicle could be dispatched based on a passenger’s needs (e.g., a smaller vehicle for a solo traveler). Greenblatt and Saxena studied trip-specific (i.e., right-sized) automated taxis based on the average proportion of occupants and total VMT. They concluded that trip-specific automated taxis could improve the fuel efficiency of fleets by 30−35%.[13] Wadud et al. investigated an extreme scenario in which all trips occur in optimally sized vehicles. In this scenario, solo travelers travel in single-occupant CAVs with the energy efficiency of motorcycles (half the fuel economy of a compact car), two-person groups travel in compact cars, groups of 3−4 travel in midsize vehicles, and groups of 5 or more travel in minivans. They reported that such a scenario would yield fuel savings of 45%.[12] While right-sizing 100% of vehicle trips may be an unrealistic goal, this demonstrates the high potential of CAV right-sizing for improving fuel economy and consequently reducing environmental impacts.

*3.1.3.c. ICT Equipment and Aerodynamic Shape Alteration.* Figure 2 shows a schematic view of information and communications technology (ICT) devices that could be added onto a generic CAV. Manufacturing ICT devices is highly carbon-intensive,[71] which increases GHG emissions associated with vehicle manufacturing. Moreover, additional ICT devices in CAVs are expected to consume more auxiliary power, which implies more operational energy use.[64] Although highly uncertain, Gawron et al. suggested that CAV subsystems and ICT equipment could increase a vehicle’s life-cycle primary energy use and GHG emissions by 3−20% because of increases in power consumption, weight, and data transmission.[48]

Furthermore, adding ICT devices, such as GPS antennae and LIDAR (light detection and ranging), could alter vehicle aerodynamics. ICT devices can create sharp edges and increase frontal projected area, both generate turbulence around the vehicle at high speeds and force the vehicle to consume more energy to maintain its performance. This could dramatically reduce CAV fuel efficiency at high speeds. There is no empirical data to evaluate how significantly add-on ICT devices affect aerodynamics and efficiency, but the magnitude of impacts can be roughly approximated using effects of roof racks on conventional vehicles. Chen and Meier reported that a roof rack can increase a passenger car’s fuel consumption by up to 25%.[72] Future CAV designs could integrate ICT equipment into the vehicle body better than the example shown in Figure 2, potentially improving aerodynamics.

*3.1.4. Platooning.* Platooning is synchronized movement of two or more vehicles trailing each other closely. Platooning reduces aerodynamic drag for following vehicles, making the whole platoon more efficient. Aerodynamic drag forces are proportional to the second power of speed, meaning that platooning is most effective in high speeds. Since platooning is practically viable for highways, adoption of this technique could yield significant fuel savings and emissions reductions. The magnitude of benefits depends on a number of platoon-specific characteristics, including cruising speed, speed variations, vehicle trailing space, vehicle shape (baseline aerodynamics), platoon size, the fraction of time spent on the highway, and the control algorithms used by the vehicles.[19,74] Vehicles in the middle of a platoon realize the largest energy efficiency gains, while gains are smaller for vehicles at the front and rear of a platoon. Longitudinal controls, sensing, and V2V communications make it possible for CAVs to safely trail each other at close distances, enabling platooning.[4] Because of the relatively slow reaction time of humans, platooning is not safe when the driver is in the loop (i.e., when driving is not fully automated).

A number of studies have experimentally shown the energy and emission effects of drag minimization by vehicle platooning.[58,75,76] Many of these experiments have focused on trucks. Given the large frontal area and high percentage of highway cruising mileages in commercial heavy-duty trucks, truck platooning would yield substantial energy savings.[77] Tsugawa reported that a 3-truck platoon traveling at 80 km/h achieves a 10% drop in energy consumption (relative to three trucks traveling conventionally) when there is a 20-m gap between trucks, and a 15% drop when the gap narrows to 5 m.[78]

For platoons containing mixed vehicle types separated by half- to full-vehicle lengths, the drag reduction is reported between 20 and 60%.[79] Wang et al. showed that a higher penetration rate of intelligent vehicles (similar to CAVs) in a tight platoon (i.e., a platoon with a very small gap between vehicles) could result in lower nitrogen oxide emissions.[58] Barth et al. projected 10−15% energy savings for platoons operating at separations of less than 4 m.[19] Similarly, Brown et al. estimated about 20% energy savings during the approximately 50% of personal vehicle travels that typically occur on highways, equating to a 10% improvement in energy efficiency overall.[14]

Platooning in dedicated lanes results in the highest environmental benefits. However, there are still beneficial opportunities for groups of two or more CAVs to platoon on mixed-use roads or lanes.[14] Platooning can also mitigate congestion and expand roadway capacity (discussed in section 3.2.4). Although the environmental benefits of platooning have been proven, research is needed to quantify expected benefits at various CAV penetration scenarios. Realizing benefits also requires new engineering design for safe platoon maneuvers—including exiting a platoon and merging—for various vehicle types.

### 3.2. Transportation System Level.

Large-scale penetration of CAVs will change transportation network loads[80] and consequently environmental impacts associated with the transportation system. The net result is difficult to predict, particularly for different levels of CAV market penetration. Major mechanisms by which CAVs affect environmental impacts of the transportation system include changing travel cost, changing mobility services, and influencing congestion and roadway effective capacity.

#### *3.2.1. Travel-Cost Implications.*

CAVs allow passengers who would normally be driving to instead occupy travel time with a variety of activities, such as working, reading, watching movies, or eating. By substituting driving for productive or leisurely activities, the perceived cost of in-vehicle time (often called "value-of-travel time" (VOTT) or "willingness to pay" to save travel time) could be diminished. Moreover, eliminating the labor cost of human drivers in transportation services reduces direct travel cost and hence expands access to transportation services for lower-income individuals and households. This socioeconomic benefit could have accompanying environmental benefits if transportation services become cheap enough that lower-incomes substitute transportation services for private vehicles and if transportation services employ energy-efficient CAVs, since lower-income households tend to drive less efficient vehicles.[81] However, lowered travel cost is expected to increase travel demand, a key effect that could yield undesired consequences.

Many studies have attempted to analyze the general cost of travel in CAVs. It is found that SAEVs could profitably reduce fees charged to passengers by up to 80% compared with a ride-on-demand trip today, a drop that would make SAEVs price-competitive with mass transit.[82] Chen and Kockelman suggested that the total cost of charging infrastructure, fleet ownership, and energy for SAEVs ranges from \$0.42 to \$0.49 per occupied mile of travel,[33] which is substantially lower than current costs of traveling in taxis or ride-hailing services. Greenblatt and Saxena showed per-mile operation cost of high-VMT SAEVs are about one-fifth of typical per-mile taxi fares.[13] Lu et al. found that automated taxis (electric and conventional) could reduce daily commute costs by over 40% but increase total transportation-related energy consumption and emissions in Ann Arbor, MI.[26]

Bosch et al. provided a more conservative estimate, indicating that shared and pooled CAV travel is likely to be only slightly less expensive than personal vehicle travel in terms of per-passenger-kilometer cost. This is because of the higher capital cost and cleaning and maintenance needs of shared fleets. They also asserted that private ownership of CAVs might be cost-competitive, despite the general assumption that SAV-based travel is cheaper than private CAV-based travel.[83] Wadud analyzed the total cost of ownership for CAVs and implications for different levels of income. The study concludes that full automation in personal vehicles offers substantial benefits for the wealthy who have a higher value of time and drive more frequently. In contrast, full automation in commercial taxis is beneficial to all income levels.[57]

The upshot is that while reducing travel costs is a positive externality likely to improve access to affordable travel options, transit equity, and consumer welfare, it may result in higher levels of energy consumption and environmental impacts at the transportation system level due to rebound effects (discussed further in section 3.4). This may offset some efficiency benefits of CAVs at the vehicle level. Moreover, the lower cost of CAV travel may discourage travelers from ride-sharing, since the cost savings associated with SAVs over private CAVs may not be substantial enough to be worth the extra hassle and reduced privacy.[83]

#### *3.2.2. Changed Mobility Services.*

CAVs could reshape mobility services by promoting shared mobility and interacting with mass transit, as discussed below.

##### *3.2.2.a. Shared Mobility.*

Large-scale penetration of CAVs has the potential to shift the transportation system from relying on privately owned vehicles to a new system relying primarily on on-demand shared mobility services,[32] commonly known as "Mobility as a Service" (MaaS).[81] Shared mobility is an effective way to reduce VMT by combining trips that are temporally and spatially similar, generating many benefits including efficiency improvements, fleet downsizing, conges-tion reduction, energy conservation, and emissions alleviation. These benefits are maximized by combining shared mobility and vehicle automation.

CAVs can help boost car-sharing by improving user experience, avoiding vehicle unavailability and inaccessibility.[84] Kang et al. proposed a system-optimization framework for automated EV sharing and suggested higher profitability and lower emissions per passenger-mile of operation compared to conventional car-sharing services.[23] CAVs can also help improve ride-sharing efficiency. Ride-sharing is intended to improve vehicle occupancy by filling empty seats in vehicles with riders on similar routes. Compared to car-sharing, ride-sharing is more dynamic and reliant on real-time matching.[85] Ride-sharing is particularly suited to CAV fleets that can continuously reroute based on real-time ride requests. Since SAVs have not yet been tested in the real world, most studies examining the topic have attempted to simulate the impact of implementing a SAV fleet in a specified area using agent-based models rather than empirical data.[26,29,31,86]

There are several ways in which combining shared mobility with CAVs can reduce travel costs. First, shared mobility systems spread ownership costs (i.e., depreciation, financing, insurance, registration, and taxes) and operational costs across a large user base.[81] Second, the shift from personally owned vehicles to on-demand SAVs could maximize capacity utilization and improve vehicle utilization rate. For instance, the average daily parking time of current private vehicles is more than 90%, with the average daily driving of approximately 30 miles.[14] However, a SAV could travel more than 200 miles and complete around 20 trips per day on average, which translates into a more efficient vehicle utilization.[26,31,87] Third, high vehicle occupancy decreases energy use per passenger-mile-traveled, which reduces the fuel cost for each passenger. Finally, a transportation system that integrates SAVs can benefit from the efficiency of centralized planning. Decisions made at fleet management businesses are more likely to consider fuel costs and prioritize efficiency compared to individual vehicle owners, who are likely to prioritize the utility of their vehicles.[37]

A number of studies find similar or lower costs for SAVs compared to current taxi services which on average cost approximately \$0.80 to \$5.75 per passenger-mile.[26,32,37,43] Fagnant and Kockelman conducted various simulations and found that the per-mile cost of a SAV fleet is around \$1.00.[29] Chen at al. estimated that the per-mile cost of a SAEV fleet ranges from \$0.75 to \$1.00.[31] Bauer et al. reported the range of \$0.29−0.61 per revenue mile of SAEV operation as a replacement for NYC taxis, which is an order of magnitude lower than the cost of present-day service.[43]

SAVs also make it possible to decrease total fleet size and/or number of vehicles operating at a given time. This yields traffic and environmental benefits by reducing congestion, increasing highway capacity, and lowering emissions (further discussed in section 3.2.3). Alonso-Mora et al. showed that introducing high-capacity CAVs with dynamic ride-sharing could substantially downsize the NYC taxi fleet. They demonstrated that using ten-passenger-capacity CAVs could serve 98% of the travel demand with a mean waiting time of 2.8 min, while shrinking the taxi fleet to 15% of its present size.[24] SAVs also make it possible to decrease the size of the private vehicle fleet while meeting current travel demand. Studies showed that one SAV could feasibly replace anywhere from 5 to 14 private vehicles.[26,29,31,88,89] The replacement rate of SAEVs depends on battery capacity and charger availability.[33,87] SAEVs have lower replacement rates than SAVs because SAEVs need to be charged, a process that takes longer than conventional refueling. Hence more SAEVs than SAVs are needed to meet the same travel demand, since there must be sufficient SAEVs available to provide service while other SAEVs are charging.[87]

*3.2.2.b. Interaction with Mass Transit.* Besides providing door-to-door mobility service, CAVs could interact with other transportation modes, such as public transit. CAVs offer a convenient option for short, frequent trips, such as traveling from subway stops and bus stations to work or home. Integrating CAVs with mass transit therefore provide a promising solution to the "first/last-mile" problem, making mass transit more convenient which can in turn reduce vehicular travel.[90] Moorthy et al. found that traveling via public transit with CAV last-mile service could reduce energy consumption by up to 37% compared to traveling with personal vehicle.[55] If automation could be expanded to buses and rail, labor cost savings could be passed onto passengers via lower trip fares, thereby improving the competitiveness of mass transit. CAV services could also be used by transit agencies in public-private partnerships to supplement or replace costly services such as low-ridership bus lines or paratransit.[6]

In contrast, CAV adoption could decrease the number of mass transit users since inexpensive CAVs could compete with transit systems. Similarly, low-cost, CAV-enabled shared mobility may result in less ridership for mass transit. Less revenue for mass transit has a disproportionate impact on low-income population, since low-income population tends to rely on transit more heavily than higher-income population.[81] Further studies are needed to quantify the likely impact of CAVs in this regard.

*3.2.3. Vehicle Utilization.* In a CAV-enabled transportation system, more people would likely be willing to travel extended routes by car[42,91] since the burden of driving is eliminated. Given that CAVs, unlike human drivers, do not need to rest, their deployment is likely to increase vehicle utilization and/or vehicle-hours-traveled. This translates to increased total VMT, energy use, and emission (further discussed in section 3.4.1).

Some studies have also found that replacing personal vehicles with SAVs will generate unoccupied VMT (e.g., as a vehicle returns to its origin after dropping off passengers), leading to higher total VMT at the transportation system level. The extent to which total system-wide VMT will change largely depends on how frequently trips are shared.[26] Fagnant and Kockelman found that if rides are never shared, a SAV-only fleet will generate 8.7% more VMT compared to a private-vehicle-only fleet, but allowing dynamic ride-sharing in a SAV fleet reduces this figure to 4.5%.[88] Similarly, Zhang et al. showed that a pooling SAV fleet generates 4.7% less VMT than a nonpooling SAV fleet.[89] Taking realistic traffic flows into account, Levin et al. reported that empty repositioning trips made by SAVs without dynamic ride-sharing increase congestion and travel time by 3−20%.[92] SAEVs could also drive to remote locations for charging, resulting in higher VMT. Loeb et al. estimated that travel to charging stations accounts for about 32% of unoccupied VMT in SAEV fleets.[87] Zhang et al. suggested that private CAVs can also generate unoccupied VMT if they reduce the number of household vehicles while maintaining the current travel patterns. For instance, a privately owned CAV could take one member of household to work, return home unoccupied, and then take another member to school. This study estimated that such relocation could increase total VMT for privately owned vehicles by around 30%.[62]

It is possible that the adverse environmental effects of CAV-related VMT increase at the transportation system level could be offset by CAV-related efficiency gains at the vehicle level (section 3.1).[17,42] It is important to note that most studies on CAV utilization assume a low SAV adoption rate (around 10%).[87−89] Increasing SAV penetration is likely to save system-wide VMT compared to a private-vehicle-only fleet, since more opportunity is available to consolidate sharable VMT and reduce unoccupied travel of SAVs due to the reduced need of vehicle relocation between trips. Moreover, some argue that CAVs could help avoid unnecessary "cruising for parking" VMT through automated navigation and parking.[14] Increasing the waiting time deemed tolerable for automated taxis would further reduce total VMT and required fleet size.[26,43]

*3.2.4. Congestion and Road Capacity.* Traffic congestion and idling contribute to additional energy use and emissions. Every new vehicle on the road uses capacity and increases congestion. Constructing new roads and lanes is one way to alleviate congestion. However, research has demonstrated that induced vehicle travel (shifts from other modes, longer trips and new vehicle trips) often consumes a significant portion of new capacity added to congested roads.[93] Alternative, arguably more sustainable options are to encourage mixed-land use and promote ride-sharing. Since SAVs can replace conventional cars at a higher rate and increase vehicle utilization efficiency (both leading to fleet downsizing), they can reduce congestion without adding road capacity. CAVs can expand effective road capacity by not only decreasing the number of vehicles on road, but also right-sizing vehicles.[12] Vehicle right-sizing will substantially reduce the fraction of fleets composed of large vehicles traveling frequently with few passengers.[13,37] While the impacts of vehicle right-sizing and fleet downsizing on improving road capacity are intuitive and frequently mentioned, quantitative estimates are missing from the literature.

Traffic jams resulting from collisions can cause congestion too. The safety improvements of CAVs is estimated to reduce congestion by 4.5% through decreasing crash frequency.[42] CAV technology can also alleviate congestion and improve effective roadway capacity by allowing vehicles to safely reduce following distance (headway), use existing lanes and intersections more efficiently by maintaining shorter distances between vehicles,[80,94] travel in coordinated platoons, take routes that avoid traffic jams and low-speed zones,[14] and also dampen stop-and-go traffic waves.[28] Another benefit is that CAVs can operate on a flat speed range 30−70 MPH on arterial roadways, which helps reduce traffic congestion.[30] Finally, CAV technology enables vehicles to synchronize movement with traffic signals, which reduces frequent acceleration and deceleration at intersections (also discussed in section 3.1). Some studies have suggested that it may be ultimately possible to achieve "signal-free" transportation systems under high CAV penetration.[54,80] Realizing such systems require major infrastructure overhauls as well as technical solutions to address pedestrian movement.

Multiple studies consider the aforementioned points in their simulations. Auld et al. applied an integrated model to analyze the impact of different market penetrations of CAVs on performance of the transportation network and changes in mobility patterns for the Chicago region. They presented a scenario in which CAVs could yield an 80% increase in road capacity with only 4% induced additional VMT.[25] Li et al. found high-CAV-penetration scenarios can reduce carbon monoxide, $PM_{2.5}$, and energy consumption in urban areas by up to 15% because of reduced congestion or increased road capacity.[36]

It is possible that vehicle automation could increase travel demand, thereby offsetting decongestion benefits. Zakharenko held that the impact of induced travel is unlikely to be very large, since CAVs and SAVs are expected to operate far more efficiently even if their utilization increases.[60] Additional research is needed to estimate the expected effects of increased travel demand on road congestion and capacity at various CAV penetration levels.

**3.3. Urban System Level.** Today's urban systems have largely been designed to accommodate privately owned and driven cars. CAVs can reshape urban systems and infrastructure in several ways. Because of improved communications, CAVs may require less infrastructure, such as traffic lights, parking lots, and road lanes. CAVs can also resolve charging-infrastructure challenges, thereby supporting vehicle electrification. However, CAVs will require additional ICT supports, though such supports could potentially be integrated into existing street lights, signs, and other transportation infrastructure. There are also concerns that CAVs could encourage suburbanism and urban sprawl.[60]

*3.3.1. Infrastructure Implications.* Deployment of CAVs will revolutionize the conventional urban infrastructure. V2I and higher safety capabilities of CAVs may render much existing infrastructure obsolete, while requiring new types to be installed. The net environmental impacts of CAV-related changes in infrastructure are largely unknown. The following sections summarize what is known and highlight priority research areas.

*3.3.1.a. Existing Infrastructure (Lighting and Traffic Signals).* Because CAVs may not need lighting for navigation or signaling, it may be possible to save energy by reducing the number or utilization of road lights and traffic lights. There is no direct data on the energy demand of road lighting and traffic signals in the U.S. The EIA estimates that in 2015, about 404 TWh of electricity was used for residential and commercial lighting.[95] This was about 15% of the total electricity consumed by both of these sectors and about 10% of total U.S. electricity consumption. On the basis pf the Department of Energy's report on U.S. Lighting Market Characterization,[96] we estimate that highway lighting (excluding traffic signals) consumes around 1% of electricity generated in the U.S. Thus, reducing road lighting by 30% would save 16.5 TWh of energy, 11 MMTs of $CO_{2eq}$, and around \$1.65 billion annually. As a comparison, in the UK, road lighting and traffic signals consume 2.5 TWh of electricity annually, representing 0.73% of total annual electricity consumption.[97]

Nevertheless, navigation is not the sole purpose of road lighting. Many passengers may not feel safe on dark roads even if CAVs can drive without risk. Some studies proposed replacing conventional road lights with intelligent and adaptive systems.[98,99] These systems could turn lights on when a CAV approaches and dim or turn lights off when the roadway is empty. V2I capabilities of CAVs facilitates such technology. Future research should examine the potential for reducing road lighting at various levels of CAV penetration from cost, maintenance, and passenger-comfort standpoints. Research should also consider different technical scenarios. For instance, the ongoing transition to light-emitting diode (LED) street lighting is increasing efficiency and so lessens the impact of eliminating lighting altogether.

*3.3.1.b. New Infrastructure Requirements.* Communication and data transmission are essential to CAV operations. CAVs depend on high frequency of information exchange for finding pick-up locations, efficient routing, and arriving safely at the final destination. All this communication and data processing requires significant computational resources and large-scale infrastructure (e.g., datacenters). The life-cycle of ICT infrastructure is energy intensive and generates a variety of environmental impacts.[71,100,101] Kolosz and Grant-Muller considered embodied emissions of roadside infrastructure and datacenters for the Automated Highway System (AHS), a system that accommodates vehicles with intelligent speed adaptation features. They reported that, despite these emissions, AHS would save an expected 280 kilotons of $CO_{2eq}$ over 15 years of operational usage in the M42 corridor, the UK's busiest highway. This is because AHS-enabled optimization of vehicles on highways reduces emissions to an extent that offsets infrastructure-related emissions.[35] More research is needed to quantify the expected net energy use and life-cycle environmental impacts of a typical datacenter for management and communications of CAV fleets.

*3.3.2. Integration of CAVs with Power Systems.* As discussed in section 3.1.2, vehicle automation and electrification are mutually reinforcing. Integrating CAVs with urban power systems can offer multiple environmental benefits.[102] Fleets of CAVs can help promote vehicle electrification by resolving challenges such as range anxiety, access to charging infrastructure, and charging time management, since connected vehicles are always aware of the availability and location of charging options.[33,46]

Automated charging infrastructure enables more efficient energy management and facilitates vehicle-grid integration and uptake of renewable electricity in transportation sector. Some prototypes of charging robotic arms and mechanisms have recently been introduced to automatically plug into EVs and control the charging process. Wireless power transfer (WPT) is a nascent technology that can complement CAVs.[103] When wireless charging is combined with CAVs, it becomes possible to automatically rotate vehicles on charge transmitter pads without human intervention. Removing this labor cost for service would make SAEVs cheaper. In addition, CAVs could navigate themselves to wireless charging spots to top up at reduced energy rates during off-peak hours. Chen et al. investigated the charging-infrastructure requirements of SAEVs and concluded that by replacing attendant-serviced charging with automated wireless charging, the operational cost of SAEV fleets drops by 20−35%.[31]

A step beyond stationary WPT is in-motion dynamic charging, in which embedded transmitters in roadways wirelessly charge vehicles as they are moving, extending maximum range or reducing the required size and cost of batteries.[103] Lavrenz and Gkritza studied the automated electric highway systems (AEHS) powered by inductive charging loops embedded in the roadway and estimated that AEHS would decrease fossil-fuel energy use by more than 25% and emissions by up to 27%.[52]

An interesting potential use of electric CAVs is as mobile energy storage units for excess electricity generated by utility-scale power plants. Under such a scheme, CAVs would automatically charge (take up power) at off-peak hours when rates and demand are low and discharge (release power) back to the grid during peak hours or in case of an electricity storage. Such bidirectional power transfer could be managed by CAV communications with the power grid and would be particularly useful in facilitating increased penetration of intermittent renewable energy like wind and solar. One caveat is that frequent charging and discharging of vehicle batteries might result in accelerated battery degradation.[103] Another is that some consumers might be reluctant to allow their privately owned vehicles to be leveraged in such a manner, even if financial incentives were provided.[104]

It is also important to note that the charging patterns of SAEVs and privately owned CAVs might be very different from charging patterns of human-driven EVs including privately owned EVs as well as EVs owned by transportation network companies.[23,43] SAEVs might need more frequent charging given their higher utilization rate (discussed in section 3.2.2). The impacts of different charging patterns on the grid and associated environmental consequences are uncertain and require further investigation.

*3.3.3. Land Use.* Because CAVs can navigate themselves to and from dedicated parking areas, increased CAV penetration reduces the need for parking located close to all destinations and hence the total amount of space needed for parking overall.[61] Nourinejad et al. noted that CAVs can park in much tighter spaces, reducing needed parking space by what they found to be an average of 67%.[105] Similarly, Zhang and Guhathakurta suggested that SAVs could reduce parking land by 4.5% in Atlanta at penetration as low as 5%.[34] Avoiding the construction of new parking could also have substantial environmental benefits. Chester et al. reported that parking construction can add 6−23 g $CO_{2eq}$ per passenger-kilometer-traveled to the total life-cycle emissions of a vehicle (typically about 230 to 380 g $CO_{2eq}$) and increase sulfur dioxide and $PM_{10}$ emissions by 24−89%.[106]

Eliminating obsolete transportation infrastructure could enable denser development in urban areas.[14] However, there are concerns that CAVs could encourage suburbanism and urban sprawl, especially for people with lower perceived values of travel time. According to Bansal et al., deployment of CAVs will likely result in long-term shifts in which people choose to relocate their homes.[38] Large families or those who tend to take advantage of lower land prices in suburbs may use CAVs to reside further from urban cores.[107] Zakharenko provided a comprehensive overview of how urban areas could be altered by CAV deployment.[60] Such qualitative discussion is common in the literature, but more quantitative analyses are needed to inform land-use policies and urban planning.

**3.4. Society Level.** The potential environmental implications of vehicle automation are the largest at the society level, but the magnitude and direction of influences are highly uncertain. One key factor is the effect that CAVs will have on public perception of mobility. For many decades, cars have been used to make a statement about individual personalities and values and often to flaunt wealth. Moreover, automakers are strongly motivated to maintain the current emotional connection of consumers to their cars,[83] unless they adopt new business models. Public perception of shared and automated driving versus private, human driving will affect the extent to which people are willing to give up private vehicles in favor of CAVs, how car manufacturers develop and market CAVs, tax and insurance policies, and infrastructure investments. Given that CAVs are not yet commercially available, assessing public opinion and consumer choice on market penetration is challenging.[39,74]

A number of surveys and questionnaires have quantified early public perception of various CAV technologies. Bansal et al. surveyed Texan families and found that more than 80% of respondents would increase vehicle utilization under a CAV paradigm.[107] König and Neumayr provided empirical evidence on mental barriers and resistance toward CAVs and suggested that people are ready and interested in riding with CAVs but not willing to buy one.[51] Kyriakidis et al. surveyed 5000 people on their acceptance of, concerns about, and willingness to buy partially, highly, and fully automated vehicles. Results indicate that respondents who are willing to pay more for fully automated vehicles are likely to have higher annual VMT and utilization rates.[41] Wadud et al.[12] and Anderson et al.[6] stated that the utilization of privately owned CAVs and induced travel demand are expected to have game-changing influence on their energy consumption and environmental impacts.

A significant negative externality of CAVs will be reduction in demand for human labor in services such as taxis, trucking, and delivery, thus potentially unemployment for many service drivers. But CAVs are expected to generate new and high-quality jobs in hardware/software technologies and in fleet management and services.

*3.4.1. Behavioral Response and Travel Pattern Shift.* The convenience, accessibility, and lower travel cost of CAVs may shift travel patterns and induce higher travel demand, mainly due to travel behavior changes. As discussed in section 3.2.1, automated driving would allow people to participate in other pursuits during their trips, lowering the perceived cost of travel and increasing acceptable commute distance and time.[17,38,42] People may prefer SAVs and SAEVs to public transit if costs are comparable, since the former options provide door-to-door service. Similarly, for short trips, people may substitute CAVs for other—often more sustainable and active—modes such as walking or cycling. It is also possible that travelers consider rechaining their trip needs (shopping, recreational, commute, errands, etc.) once they have access to CAV technology. Overall, CAVs have the potential to replace not only private vehicles but many other types of transportation.

CAVs could also unlock additional travel demand from people who have unmet travel needs and previously cannot or choose not to drive (e.g., the elderly, the young, unlicensed individuals, and people with driving-restrictive medical conditions or disabilities). CAVs can provide door-to-door mobility service for these populations that is cheaper and more convenient than current options like paratransit or taxis. Expanded mobility for currently underserved population is highly desired from an equity and ethical standpoint but is likely to increase trip frequency—especially in suburban, vehicle-dependent areas.[17] Harper et al. estimated that the increase in travel demand from travel-restricted population could be as much as an additional 14% VMT (equivalent to 295 billion miles) per year in the U.S.[40]

Increased travel demand associated with CAVs represents a type of "rebound effect." In the energy economics, rebound effects describe the percentage of energy savings from a new, energy-efficient technology that are offset by increased use of that technology.[108] Similarly, efficiency gains from CAV technology at the vehicle level may induce additional travel demand and consequently offset environmental benefits at the society level. Such rebound effects can cause discrepancies between predicted and realized net impacts of CAVs and other transportation innovations.[109]

For CAVs, the rebound effect is one of the mechanisms connecting different system levels. Milakis et al. presented a ripple model to conceptualize rebound effects in societal aspects of automated driving.[27] Wadud et al. used a simple approach to employ rebound effects from generalized cost of travel as a multiplier of CAV travel activity by simulating a range of literature-driven travel elasticities.[12] In short, it is widely accepted that rebound effects could offset environmental benefits of CAVs, but there is significant uncertainty about the extent. Considering the importance of this issue for the environment as well as for transportation and infrastructure planning, additional effort to model and quantify CAV-related rebound effects is urgently needed.

*3.4.2. Shared Consumption.* Public opinion on private vehicle use and social norms over vehicle ownership may change along with the introduction of shared mobility in the transportation sector.[32,110] CAVs can help change public perception of shared consumption by facilitating and promoting shared mobility.[111] The millennial generation has already shown different transportation preferences and opinions compared to prior generations.[107,110,111] We speculate that this shift might be extended to other types of goods and services. In a society where shared consumption is mainstream, desire for product ownership will be reduced, which will reduce environmental impacts associated with product life-cycles. CAV-facilitated shared mobility can support this change from a technological perspective, but questions remain as to adoption behaviors and public acceptance. The literature does not yet show what future travelers will want from their transportation systems.

*3.4.3. Transformation of Other Sectors.* Widespread deployment of CAVs may also influence other transportation industries, such as aviation and rail. Given the lower cost of CAV travel, certain groups of users may choose to take longer trips using road transportation rather than aviation or rail. This is environmentally significant, as aviation and rail tend to have lower marginal energy use and emissions on a per-passenger-mile-traveled basis compared to low- or single-occupancy vehicles.[16] Both intercity rail (56.1 passenger-miles per gasoline-gallon equivalent (GGE)) and airlines (50.0 passenger-miles per GGE) have higher energy efficiency compared to passenger vehicles (38.9 passenger-miles per GGE).[112] LaMondia et al. studied the impact of CAVs on long-distance travel choices by analyzing travel surveys, and concluded that CAVs could displace 25−35% of demand for air travel for trips of 500 miles or more.[113] The environmental impact of this shift could be mitigated if intercity CAV travels were mostly through larger shared vehicles such as autonomous buses.

CAVs are also likely to affect a variety of transport-intensive sectors and services. For instance, CAVs could serve as mobile overnight sleeping compartments, decreasing demand for hotels for long-distance trips.[91] Sectors that heavily utilize freight transportation—online retail, the food industry,[50] etc.—will likely benefit from the emergence of CAVs. The environmental impacts of CAV adoption and utilization in these sectors are likely significant, but little is known.[50] More research is needed to measure these broader impacts and inform relevant policymaking.

*3.4.4. Workforce Impacts.* Vehicle automation will render many jobs obsolete, specifically in labor-intensive transportation services such as freight trucking, public transit, and taxi driving.[27,42] The U.S. Department of Commerce estimates that 15.5 million U.S. workers are employed in occupations that could be affected by the introduction of automated vehicles.[114] Unemployment has attendant economic and social consequences. These include altered consumption patterns (usually moving toward less sustainable commodities and services), as well as adverse physical and mental health effects.[45] Both these consequences have environmental relevance as consumption pattern changes drive changes in supply chain and associated environmental impacts. It should be noted that CAV-related job losses will occur gradually in most cases. For instance, early automated trucks will still require human drivers to assist with loading and unloading, navigation, fueling, and maintenance. Over time, though, retraining the workforce and alternative job opportunities will be needed to ensure sustainable CAV adoption and mitigate adverse outcomes.[50] One option is to help workers in transportation-related jobs transition to sectors that are likely to expand as CAV penetration grows. These sectors include but not limited to hardware and software development, fleet management, and concierge services.

**3.5. Summary of Environmental Impacts of CAVs.** Our review shows that due to the complexity and interdependence of higher levels of interactions, the uncertainty of CAV-related environmental impacts increases as the impact scope broadens. Most studies related to energy and environmental impacts of CAVs have tried to identify effect bounds and speculate on system-level impacts. Collectively, these studies confirm that CAV technology has the potential to deliver large environmental benefits, but realizing this potential highly depends on deployment strategies and consumer behavior. The greatest energy and environmental impacts will not stem from CAV technology directly, but from CAV-facilitated transformations at all system levels.

At the vehicle level, CAV technology can significantly enhance efficiency. Considerable fuel savings and emission reduction can be achieved through CAV design oriented toward energy efficiency. Studies reviewed in this paper report vehicle-level fuel savings ranging between 2% and 25% and occasionally as high as 40%. Integrating CAV technology and vehicle electrification can considerably improve the economics and attractiveness of transportation decarbonization. Higher CAV penetration could further alleviate negative environmental impacts of road transportation through large-scale, connected eco-driving. However, the net effect of CAV technology on energy consumption and emissions in the long term remains uncertain and depends on other levels of interactions with the environment.

At the transportation system level, CAV-related environmental benefits derive from optimization of fleet operations, improved traffic behavior, more efficient vehicle utilization, and the provision of shared mobility services. Specifically, shared mobility and CAV technology have significant mutual reinforcing effects.

At the urban system level, CAVs could reshape cities by changing land-use patterns and transportation infrastructure needs. For instance, street lighting and traffic signals could become less necessary or obsolete under a CAV paradigm, resulting in energy savings. However, CAVs could encourage urban sprawl and shifting to peripheral zones with longer commutes. CAVs also require communications with large-scale datacenters, which are generally energy intensive. At the same time, CAVs can facilitate integration of EVs and charging infrastructure into power grids. These urban-level mechanisms might not deliver significant net environmental benefits without high penetration of CAV technology.

While long-term net environmental impacts of CAVs at the vehicle, transportation system, and urban system levels seem promisingly positive, the lower cost of travel and induced demand at the society level is likely to encourage greater vehicle utilization and VMT. Most studies reviewed in this paper assume current travel patterns, vehicle ownership models, and vehicle utilization without considering realistic behavioral changes resulted from increased CAV penetration. Society-level impacts of CAVs will undoubtedly be profound, but significant uncertainties exist about behavioral changes, making it very difficult to project the actual energy and environmental impacts.

The synergetic effects of vehicle automation, electrification, right-sizing, and shared mobility are likely to be more significant than any one isolated mechanism. Hence, these synergies should be the focus of future research efforts. Fulton et al. projected that the combination of these technologies could cut global energy use by more than 70% and reduce $CO_2$ emissions from urban passengers by more than 80% by 2050.[47] They further estimated that the combination of these technologies could reduce costs of vehicles, infrastructure, and operations in the transportation sector by more than 40%, achieving savings approaching $5 trillion annually compared to the business-as-usual case.

To ensure truly sustainable uptake and adoption of CAV technology, transportation systems must be more energy efficient, facilitate emissions reduction, mitigate local air pollution, and address public health concerns. At the same time, strategic development and deployment of CAV technology are necessary to control overall travel demand and congestion.

## 4. PRIORITY RESEARCH NEEDS

On the basis of our review of the literature, we recommend the following four principles for improving research on the energy, environmental, and sustainability implications of CAVs:

I. **Where possible, transition to empirical, data-based analysis of CAV impacts and revisit assumptions.** The novelty of CAV technology and lack of data means that analysis of CAV impacts has, to date, been largely speculative and qualitative. Moreover, many analyses are based on oversimplified or unrealistic assumptions. Researchers should strive to increase the rigor of CAV studies as more data and higher fidelity models become available.

II. **Improve models by more accurately characterizing CAV impacts and better capturing uncertainty.** Most analyses have assumed the mechanisms by which CAVs impact the environment are independent of one another. This assumption frequently leads to underestimation or overestimation of aggregate impacts. Furthermore, models should better reflect the true nature of CAV impacts. For instance, many studies fail to distinguish between general trends of energy efficiency improvement in vehicles and additional benefits that are solely enabled by CAV attributes. It is also necessary to quantify the upper and lower bounds of impacts and incorporate these bounds into models to better capture and characterize uncertainty.

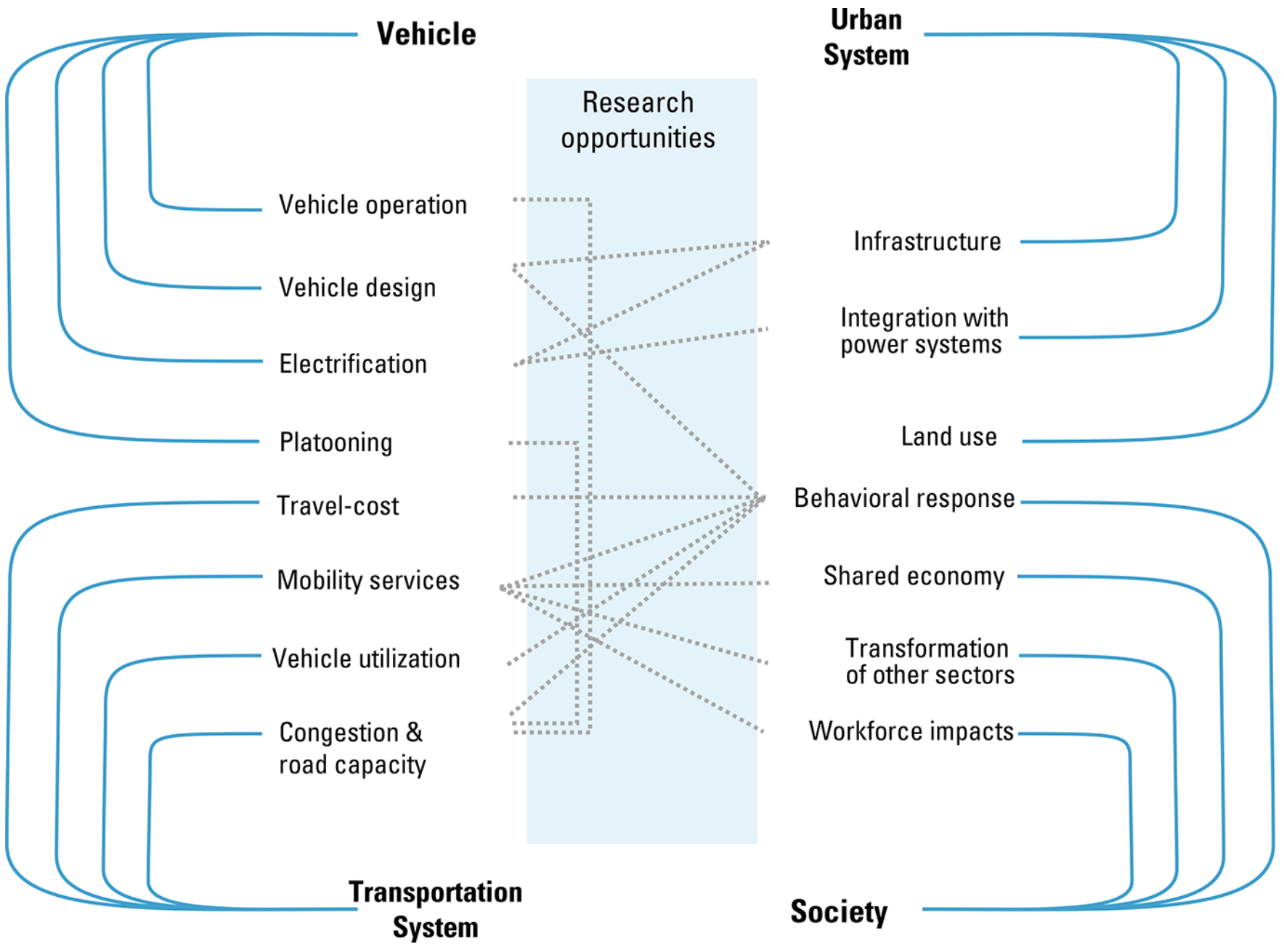

**Figure 3.** Interactions and linkages between system levels that entail energy, environmental, and sustainability impacts. The linkages are illustrative and not necessarily exhaustive.

III. **Place more effort on understanding the effects of different CAV technologies and market scenarios on consumer behavior and travel patterns.** Although improvements in CAV efficiency at the vehicle level should not be overlooked, the largest environmental impacts are likely to depend on consumer behavior and travel patterns: that is, when, where, how often, and how much consumers travel with CAVs.

IV. **Integrate analysis and modeling across different system levels.** There is a need for deeper investigation on how mechanisms at each level reinforce or undermine each other. Figure 3 illustrates interactions and linkages across the four system levels identified in this review that are likely to have substantial energy, environmental, and sustainability implications. The trade-offs between interactions and linkages are largely unexplored and merit further research.

We also recommend prioritizing research on four specific topics: CAV design and testing, development of CAV-specific models and tools, investigation of behavioral phenomena associated with CAV sharing and adoption, and assessment of policy needs and opportunities. Each of these is discussed in further detail below.

**4.1. CAV Design and Testing.** The evolution of vehicle design is a major source of uncertainty for the environmental performance of CAVs. There is a gap in the literature regarding which factors should drive the vehicle design optimization and decision-making protocols that will affect CAV-related energy consumption and emissions. Conventional life-cycle assessment (LCA) can be used to characterize the first-order impacts of various design protocols and provide insights that can improve sustainability of early CAV designs. However, for more radical and complex designs (including vehicle right-sizing and safety-enabled light-weighting), more sophisticated sustainability assessments are needed. Studies should be conducted to characterize environmental benefits of different CAV designs under different real-world scenarios and particularly under different levels of societal CAV acceptance.

Another priority should be quantifying energy efficiency improvements actually achieved by early commercial designs. Proving grounds and test facilities are needed to demonstrate that theoretical CAV efficiencies can be practically achieved. Providing researchers with real-world data from on-board diagnostics (of current prototypes) can help identify best practices and designs. Results can then be used to improve real-world development and deployment.

Considerations need to be given in carrying out such research to avoid infringing on consumer privacy or compromise intellectual property.

**4.2. CAV-Specific Models and Tools.** CAVs will have impacts on and be affected by land use, demand, demographic changes, economic factors, fueling infrastructure, and local policies, among other factors. CAV-related changes in demand for and supply of mobility services will change loads placed on transportation networks. For instance, CAVs could improve freeway traffic flows by enabling shorter following distances between vehicles but deteriorate road congestion and effective capacity by inducing more travel. Also, current vehicle-choice models are ill-suited to incorporate numerous consumer preference variables relevant to CAV adoption. Moreover, CAVs are not yet integrated into major transportation and energy models—such as those used by the U.S. DOT, EPA, EIA, and the Intergovernmental Panel on Climate Change—for estimating future travel demand, energy use, and environmental consequences. In most existing assessment studies, various measures that can reduce demand for travel or vehicle usage and improve driving performance have been identified. However, CAVs most likely entail considerable yet uncertain rebound effects, making current predictions of future transportation demand unreliable.[15] Integrated assessment models and research support tools that incorporate environmental effects of system-level CAV attributes for various market penetrations should be developed to enable higher-quality projections of future travel trends.

**4.3. Behavioral Studies.** Scant effort has been dedicated to analyzing how consumer preference for CAV technology, vehicle ownership, and ride-sharing might evolve. This is important given that the net environmental impacts of CAVs are highly dependent on the degree to which CAVs are shared versus privately owned. Pooling and shared mobility services alleviate most adverse environmental effects of CAV technology. However, social norms may lead people to avoid sharing transportation with strangers, especially if cost differences are marginal. Research is needed to identify the factors that will affect these choices. There is a particular need to examine mixed private/shared CAV scenarios, since most studies conducted to date examine scenarios in which CAVs are either fully private or fully shared.

Further investigation is also needed into how readily consumers will adopt CAVs. Real-world data can be obtained from surveys and tests. However, surveys are probably less useful due to the novelty of CAV technology, since most respondents will not be able to provide an informed response. Novel approaches are needed to investigate if and under what circumstances people will accept CAVs and how they will use them. Creative techniques such as virtual and augmented reality might be useful in this regard. More extensive engagement—i.e., participants work with researchers to understand possible technology options and more deeply explore scenarios—could also provide deeper insight into how people actually perceive CAV technology.

**4.4. Policy Needs and Opportunities.** Governments are already playing an active role in supporting technological development of CAVs. Emphasis has been placed on safety, equity, and mobility, while scant attention has been paid to environmental implications. For example, a bipartisan group of U.S. senators recently released a set of principles for self-driving vehicle legislation as part of the American Vision for Safer Transportation through Advancement of Revolutionary Technologies (AV START) Act. These principles do not mention energy, efficiency, or emissions at all.[115] This omission is problematic, given large environmental opportunities—and risks—associated with CAV technology.

Historically, the majority of environmental policies for the transportation sector have focused on regulating tailpipe emissions. Since CAVs are likely to be more efficient and generate lower levels of emissions than conventional vehicles, limiting emissions on a per-vehicle basis is less important than considering potential environmental impacts of CAVs on a broader scale. CAVs may induce travel demand that offsets—or even eliminates—improvements in per-vehicle efficiency and emissions. It is important to develop policies that address this concern. CAVs also provide new opportunities for governance. Vehicle connectivity enables environmental policies, such as mileage charges, regulation of unoccupied travel, and dynamic emission reporting.[116] Such policies have advantages. For instance, VMT taxation is seen as less regressive—hence more equitable—and more economically efficient than fuel taxes.[117] However, collecting accurate spatial and time-of-day vehicle use may raise privacy concerns and is politically difficult to implement.

In addition to exploring CAV-specific policy options, policymakers should consider establishing CAV policy frameworks that can be adapted based on how the market and technology evolves. Several possible use cases of CAVs that would have significant external costs are not discouraged by current policy, and the most beneficial use cases are not incentivized. For example, large, personally owned, inefficient CAVs could serve the owner at significant cost to the system by driving "selfishly" (for instance cruising streets empty instead of paying for parking), and underpaying for impacts on infrastructure. It remains to be seen whether this use case will manifest in reality. But implementing mechanisms—such as dynamically pricing CAV use on a per-mile basis in congested areas or at peak times—for addressing undesired outcomes will be far easier now than once CAVs are already on the road.

Overall, robust understanding of energy, environmental, and sustainability impacts of CAV technology depends on the evolution of technology, behavioral responses, market penetration, and regulatory and policy considerations. Inclusion of all relevant factors to maximize environmental benefits and minimize adverse consequences is critical for the development of this transformational transportation technology that does not only saves lives but also improves the environment.

## ASSOCIATED CONTENT

### Supporting Information

The Supporting Information is available free of charge on the ACS Publications website at DOI: 10.1021/acs.est.8b00127.

Short description of CAV components (PDF)

## AUTHOR INFORMATION

### Corresponding Author

*E-mail: mingxu@umich.edu.

### ORCID

Hannah R. Safford: 0000-0001-9283-2602

Ming Xu: 0000-0002-7106-8390

### Notes

The authors declare no competing financial interest.

## ACKNOWLEDGMENTS

Authors thank several participants of the 2017 and 2018 Automated Vehicle Symposium (Energy and Environmental Implications of CAVs Breakout Sessions), as well as many other experts for providing helpful suggestions, insight, and feedback. The contribution of Dave Brenner for creating figures is appreciated. We also thank the anonymous reviewers, whose constructive comments substantially improved the paper.